\documentclass[apjl]{emulateapj}
\usepackage{psfig,amsfonts,amsmath,graphicx,natbib,apjfonts}
\newcommand{\gps}{\ensuremath{g_{\rm P1}}}
\newcommand{\rps}{\ensuremath{r_{\rm P1}}}
\newcommand{\ips}{\ensuremath{i_{\rm P1}}}
\newcommand{\zps}{\ensuremath{z_{\rm P1}}}
\newcommand{\yps}{\ensuremath{y_{\rm P1}}}

\newcommand{\grizy}{\gps\rps\ips\zps\yps}

\def\ra#1#2#3{#1$^{\rm h}$#2$^{\rm m}$#3$^{\rm s}$}
\def\dec#1#2#3{$#1^\circ#2'#3''$}

\def\ps1{PS1-11bam}

\def\cfa{1}
\def\stsci{2}
\def\gemini{3}
\def\ifa{4}
\def\jhu{5}
\def\qub{6}
\def\pri{7}

\begin{document}

\title{Ultra-Luminous Supernovae as a New Probe of the Interstellar
Medium in Distant Galaxies}

\author{
E.~Berger\altaffilmark{\cfa},
R.~Chornock\altaffilmark{\cfa},
R.~Lunnan\altaffilmark{\cfa},
R.~Foley\altaffilmark{\cfa},
I.~Czekala\altaffilmark{\cfa},
A.~Rest\altaffilmark{\stsci},
C.~Leibler\altaffilmark{\cfa},
A.~M.~Soderberg\altaffilmark{\cfa},
K.~Roth\altaffilmark{\gemini},
G.~Narayan\altaffilmark{\cfa},
M.~E.~Huber\altaffilmark{\ifa},
D.~Milisavljevic\altaffilmark{\cfa},
N.~E.~Sanders\altaffilmark{\cfa},
M.~Drout\altaffilmark{\cfa},
R.~Margutti\altaffilmark{\cfa},
R.~P.~Kirshner\altaffilmark{\cfa},
G.~H.~Marion\altaffilmark{\cfa},
P.~J.~Challis\altaffilmark{\cfa},
A.~G.~Riess\altaffilmark{\stsci,}\altaffilmark{\jhu},
S.~J.~Smartt\altaffilmark{\qub},
W.~S.~Burgett\altaffilmark{\ifa},
J.~N.~Heasley\altaffilmark{\ifa},
N.~Kaiser\altaffilmark{\ifa},
R.-P.~Kudritzki\altaffilmark{\ifa},
E.~A.~Magnier\altaffilmark{\ifa},
M.~McCrum\altaffilmark{\qub},
P.~A.~Price\altaffilmark{\pri},
K.~Smith\altaffilmark{\qub}, 
J.~L.~Tonry\altaffilmark{\ifa}, and
R.~J.~Wainscoat\altaffilmark{\ifa}
}

\altaffiltext{1}{Harvard-Smithsonian Center for Astrophysics, 60
Garden Street, Cambridge, MA 02138, USA}

\altaffiltext{2}{Space Telescope Science Institute, 3700 San Martin
Drive, Baltimore, Maryland 21218, USA}

\altaffiltext{3}{Gemini Observatory, 670 N. Aohoku Place Hilo, HI
96720, USA}

\altaffiltext{4}{Institute for Astronomy, University of Hawaii, 2680
Woodlawn Drive, Honolulu HI 96822}

\altaffiltext{5}{Department of Physics and Astronomy, Johns Hopkins
University, Baltimore, MD 21218, USA}

\altaffiltext{6}{Astrophysics Research Centre, School of Mathematics
and Physics, Queen's University Belfast, Belfast BT7 1NN, UK}

\altaffiltext{7}{Department of Astrophysical Sciences, Princeton
University, Princeton, NJ 08544, USA}

\begin{abstract} We present the Pan-STARRS1 discovery and light
curves, and follow-up MMT and Gemini spectroscopy of an ultra-luminous
supernova (ULSN; dubbed PS1-11bam) at a redshift of $z=1.566$ with a
peak brightness of $M_{\rm UV}\approx -22.3$ mag.  PS1-11bam is one of
the highest redshift spectroscopically-confirmed SNe known to date.
The spectrum is characterized by broad absorption features typical of
previous ULSNe (e.g., \ion{C}{2}, \ion{Si}{3}), and by strong and
narrow \ion{Mg}{2} and \ion{Fe}{2} absorption lines from the
interstellar medium (ISM) of the host galaxy, confirmed by an
[\ion{O}{2}]$\lambda 3727$ emission line at the same redshift.  The
equivalent widths of the \ion{Fe}{2}$\lambda 2600$ and
\ion{Mg}{2}$\lambda 2803$ lines are in the top quartile of the quasar
intervening absorption system distribution, but are weaker than those
of gamma-ray burst intrinsic absorbers (i.e., GRB host galaxies).  We
also detect the host galaxy in pre-explosion Pan-STARRS1 data and find
that its UV spectral energy distribution is best fit with a young
stellar population age of $\tau_*\approx 15-45$ Myr and a stellar mass
of $M_*\approx (1.1-2.6)\times 10^9$ M$_\odot$ (for $Z=0.05-1$
Z$_\odot$).  The star formation rate inferred from the UV continuum
and [\ion{O}{2}]$\lambda 3727$ emission line is $\approx 10$ M$_\odot$
yr$^{-1}$, higher than in any previous ULSN host.  PS1-11bam provides
the first direct demonstration that ULSNe can serve as probes of the
interstellar medium in distant galaxies.  At the present, the depth
and red sensitivity of PS1 are uniquely suited to finding such events
at cosmologically interesting redshifts ($z\sim 1-2$); the future
combination of LSST and 30-m class telescopes promises to extend this
technique to $z\sim 4$.  \end{abstract}
 

\section{Introduction}
\label{sec:intro}

Studies of the interstellar medium (ISM) in distant galaxies have
traditionally focused on two observational techniques: (i) direct
spectroscopic measurements of emission lines from the aggregate
\ion{H}{2} regions (e.g., \citealt{sgl+05,esp+06,mcm+09}); and (ii)
absorption by intervening systems in spectra of background quasars
(e.g., \citealt{ss92,wgp05}).  These approaches are complementary:
direct galaxy metallicity measurements at $z\gtrsim 2$ are challenging
due to the relative faintness of galaxies and the redshifting of
rest-frame optical emission lines into and beyond the near-IR band,
while quasar absorption studies are most effective at $z\gtrsim 2$
where Ly$\alpha$ is redshifted into the observed optical window.

In recent years, these studies have been supplemented by absorption
spectroscopy of bright gamma-ray burst (GRB) afterglows (e.g.,
\citealt{vel+04,bpc+06,pcd+07,fjp+09}), which are detectable at least
to $z\sim 9.5$ \citep{tfl+09,clf+11}.  GRBs provide a unique view of
the ISM since their massive star progenitors are embedded within star
forming regions of their hosts \citep{bkd02}.  As a result, GRBs probe
galaxies (their hosts) at a small impact parameter of $\lesssim {\rm
few}$ kpc, and thus provide a complementary view to quasars, which
tend to probe the outer halos of intervening galaxies due to the large
cross-section at large radii.  As a consequence of this critical
difference, GRBs have revealed higher neutral hydrogen and metal
column density absorbers than quasars show (e.g.,
\citealt{bpc+06,pcd+07}).

Despite this success, GRB observations are not trivial.  First, they
require wide-field $\gamma$-ray satellites with real-time
arcminute-scale localization capability (e.g., {\it Swift}).  Second,
while GRB optical/near-IR afterglows are initially more luminous than
quasars ($\langle m(1\,{\rm hr})\rangle\sim 17$ mag, $\langle
M(1\,{\rm hr})\rangle\sim -28$ mag; e.g., \citealt{kkz+10}), the
emission fades rapidly, by $3-4$ mag in one day.  Thus, the window of
opportunity for GRB absorption spectroscopy is brief.  In this
context, an alternative astrophysical source with the advantages of
GRBs (embedded massive star progenitor, large luminosity), but that
overcomes the disadvantages of GRBs (rapid fading, $\gamma$-ray
discovery) would provide a powerful probe of distant galaxies.  Normal
supernovae (SNe) are inadequate for this purpose since with $M\gtrsim
-19.5$ mag they are too faint for absorption spectroscopy at
cosmologically interesting
redshifts\footnotemark\footnotetext{Supernovae have been previously
utilized for absorption studies of the ISM in nearby galaxies (e.g.,
\citealt{wcg72,vac+87,brm+00}).} ($z\gtrsim 1$).  Moreover, the
spectral energy distributions (SEDs) of Type I SNe are strongly
suppressed at $\lambda_r\lesssim 4000$ \AA\ due to iron line
blanketing, and they are therefore poorly suited to the detection of
interstellar Ly$\alpha$ and metal UV lines in absorption.  The SEDs of
Type II SNe can extend into the UV, but they are generally less
luminous than Type I events, although some Type IIn events are
ultra-luminous (e.g., \citep{ock+07,slf+07}).

Against this backdrop, a recently-discovered class of ultra-luminous
SNe (ULSNe) with $M\approx -22$ to $-23$ mag and SEDs that extend into
the UV may provide a powerful probe of distant galaxies
\citep{qaw+07,bdt+09,psb+10,ccs+11,qkk+11}.  This potential was first
noted by \citet{qkk+11}, and further discussed by \citet{ccs+11}.
Here we present the first demonstration of this potential with
spectroscopic observations of an ULSN at $z=1.566$ discovered in the
Pan-STARRS1 Medium-Deep Survey (PS1/MDS).  This is one of the highest
redshift spectroscopically-confirmed SNe known to date.  The spectra
exhibit interstellar absorption features from \ion{Fe}{2} and
\ion{Mg}{2} with equivalent widths that are intermediate between the
populations of quasar and GRB absorbers.  We also present PS1
detections of the host galaxy in several rest-frame UV bands that
point to a substantial star formation rate, a young stellar
population, and low stellar mass that are reminiscent of GRB host
galaxies.  These results pave the way for the use of ULSNe as probes
of distant galaxies in the LSST and GSMT era.

\section{Observations} 
\label{sec:obs}

\subsection{PS1 Survey Summary}

The PS1 telescope on Haleakala is a high-etendue wide-field survey
instrument with a 1.8-m diameter primary mirror and a $3.3^\circ$
diameter field of view imaged by an array of sixty $4800\times 4800$
pixel detectors, with a pixel scale of $0.258''$
\citep{PS1_system,PS1_GPCA}.  The observations are obtained through
five broad-band filters (\grizy), with some differences relative to
the Sloan Digital Sky Survey (SDSS); the \gps\ filter extends $200$
\AA\ redward of $g_{\rm SDSS}$ to achieve greater sensitivity and
lower systematics for photometric redshifts, and the \zps\ filter
terminates at $9300$ \AA, unlike $z_{\rm SDSS}$ which is defined by
the detector response \citep{tsl+12}.  PS1 photometry is in the
``natural'' system, $m=-2.5{\rm log}(F_\nu) +m'$, with a single
zero-point adjustment ($m'$) in each band to conform to the AB
magnitude scale.  Magnitudes are interpreted as being at the top of
the atmosphere, with 1.2 airmasses of atmospheric attenuation included
in the system response function \citep{tsl+12}.

The PS1 Medium-Deep Survey (MDS) consists of 10 fields (each with a
single PS1 imager footprint) observed on a nearly nightly basis by
cycling through the five filters in $3-4$ nights to a $5\sigma$ depth
of $\sim 23.3$ mag in \gps\rps\ips\zps, and $\sim 21.7$ mag in \yps.
The MDS images are processed through the Image Processing Pipeline
(IPP; \citealt{PS1_IPP}), which includes flat-fielding
(``de-trending''), a flux-conserving warping to a sky-based image
plane, masking and artifact removal, and object detection and
photometry; transient detection using IPP photometry is carried out at
Queen's University Belfast.  Independently, difference images are
produced from the stacked nightly images by the {\tt photpipe}
pipeline \citep{rsb+05} running on the Odyssey computer cluster at
Harvard University.  The discovery and data presented here are from
the {\tt photpipe} analysis.

\subsection{Discovery and Photometric Observations of PS1-11bam}

\begin{figure}
\centering
\includegraphics[angle=0,width=3.6in]{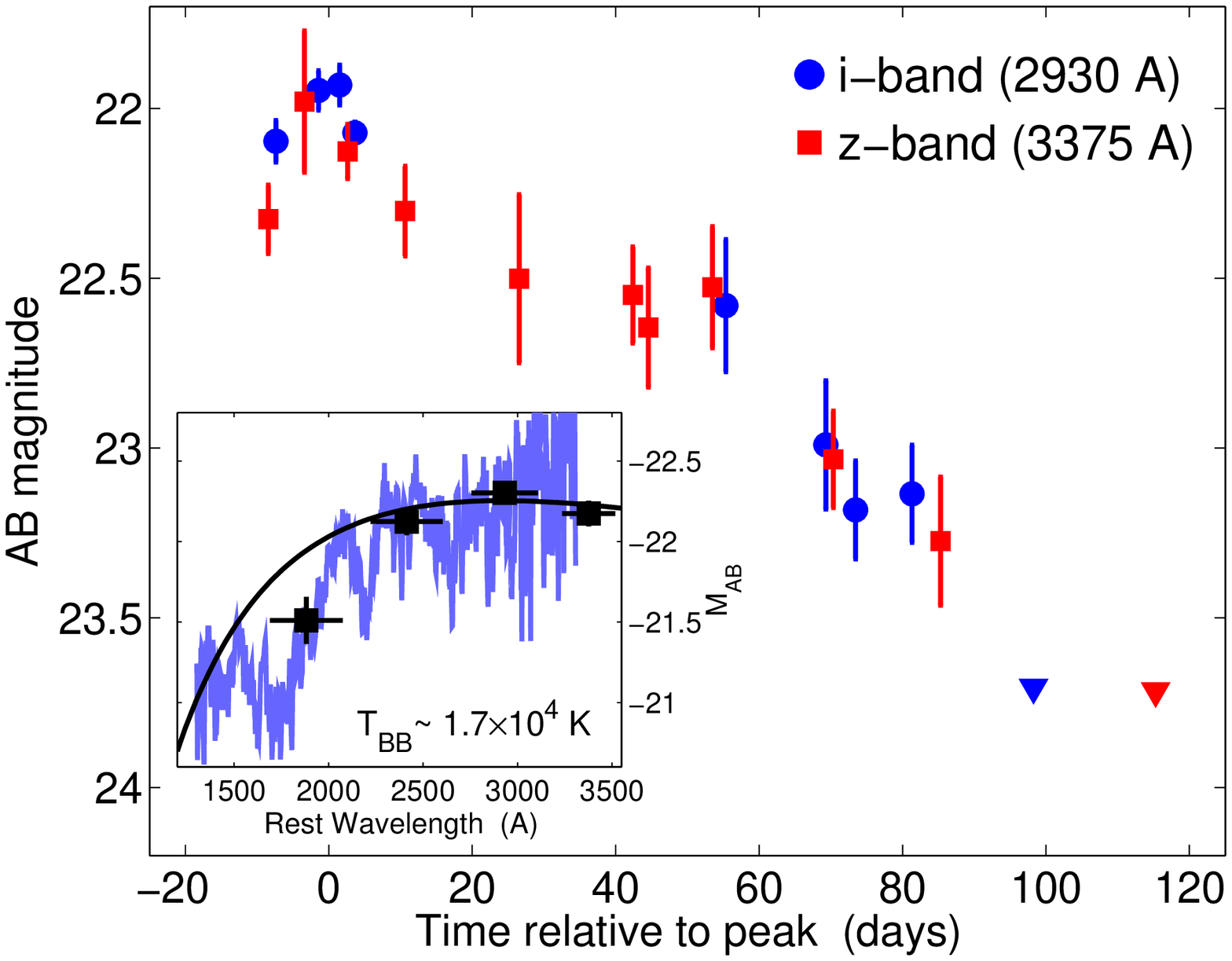}\hfill
\includegraphics[angle=0,width=3.35in]{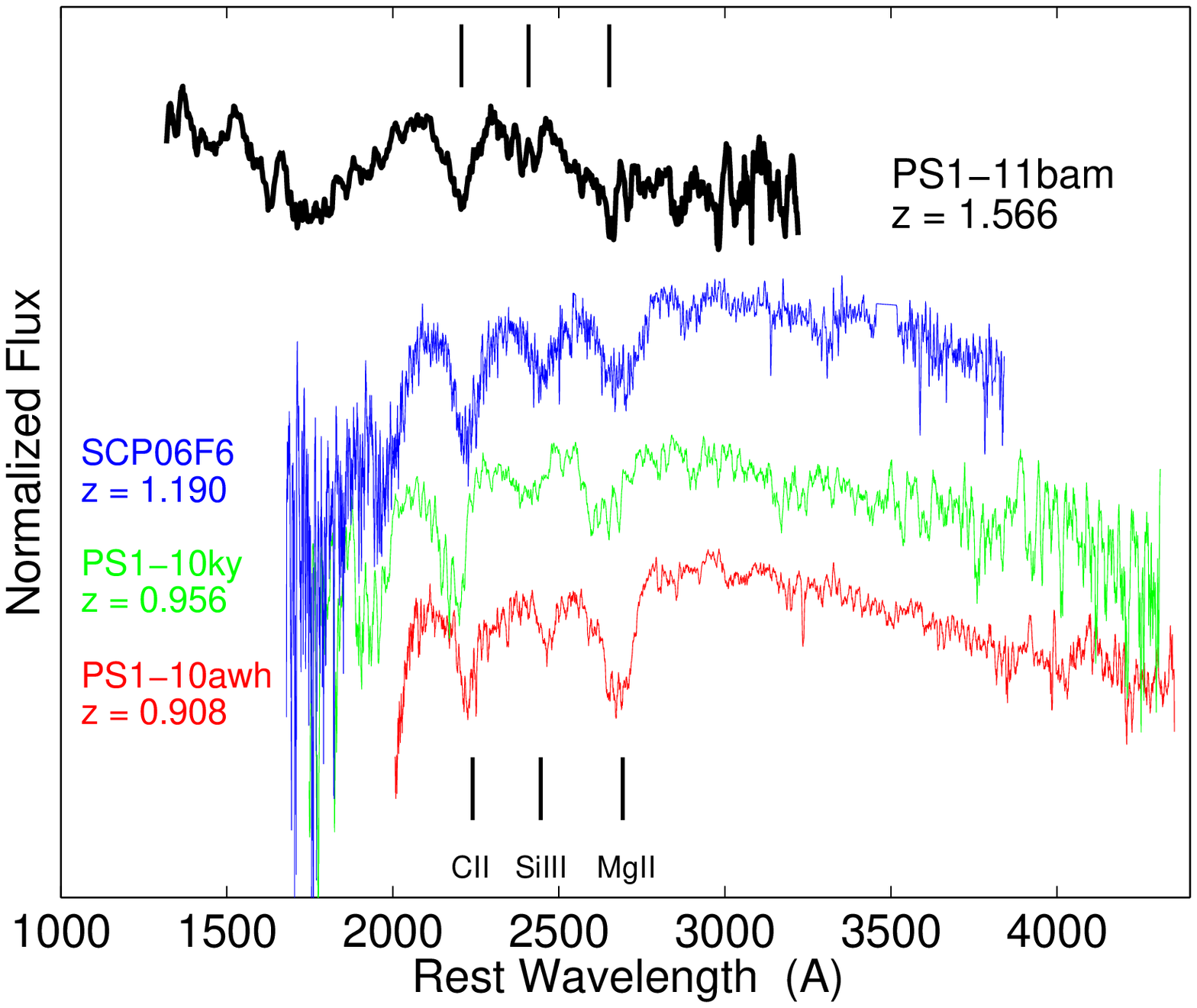}
\caption{{\it Left:} PS1 \ips\zps\ light curves of PS1-11bam in the
observer frame.  We define the peak time relative to the \ips\ data.
The transient was discovered about 10 d before maximum in the first
images of the year and faded below the PS1/MDS detection threshold at
about 80 d post-maximum.  The inset shows the SED from
\gps\rps\ips\zps\ data obtained within $\pm 1$ d of the peak, as well
as the MMT spectrum scaled by the \ips-band magnitude.  The SED
matches a blackbody spectrum with $T_{\rm BB}\approx 1.7\times 10^4$
K, with a suppression in \gps\ due to broad absorption at
$\lambda_r\lesssim 2000$ \AA.  {\it Right:} MMT spectrum of PS1-11bam
(black) exhibiting broad absorption features typical of previous ULSNe
(\ion{C}{2}, \ion{Si}{3}, and \ion{Mg}{2}).  For comparison we show
the three previous highest-redshift ULSNe: SCP06F6 at $z=1.190$ (blue;
\citealt{bdt+09}), PS1-10ky at $z=0.956$ (green; \citealt{ccs+11}),
and PS1-10awh at $z=0.908$ (red; \citealt{ccs+11}).  Flux from
PS1-11bam is detected to at least $\lambda_r\approx 1300$ \AA.
\label{fig:lcs_spec}}
\end{figure}

PS1-11bam was discovered in PS1/MDS data at RA=\ra{08}{41}{14.192},
Dec=\dec{+44}{01}{56.95} (J2000), about 10 days before maximum light,
with the time of peak corresponding to 2011 November 22 UT
(Figure~\ref{fig:lcs_spec}).  The object was detected in the first
images of the season, so the actual rise time is likely $\gtrsim 10$
d.  PS1-11bam was detected for nearly 100 d in the \ips\ and \zps\
bands, and for a shorter period of time near peak in the \gps\ and
\rps\ bands.  The peak absolute magnitude in \ips\ ($\lambda_r\approx
2930$ \AA) is $M_{\rm AB}=-22.3\pm 0.1$ mag, matching the most
luminous SNe to date (c.f., \citealt{ccs+11,qkk+11}).

The SED within $\pm 1$ d of peak brightness can be fit with a
blackbody function, with a temperature of about $1.7\times 10^4$ K
(Figure~\ref{fig:lcs_spec}).  The flux density in the \gps-band
($\lambda_r\approx 1880$ \AA) is suppressed compared to the blackbody
model due to broad absorption blueward of $\lambda_r\approx 2000$ \AA\
(\S\ref{sec:spec} and Figure~\ref{fig:lcs_spec}).

\subsection{Spectroscopy of PS1-11bam}
\label{sec:spec}

We obtained spectra of PS1-11bam with the Blue Channel spectrograph
\citep{swf89} on the MMT 6.5-m telescope on 2011 November 29 UT.  The
spectrum consisted of $3\times 1200$ s exposures obtained at an
airmass of 1.1 with a $1''$ slit aligned at the parallactic angle.
The data were processed using standard procedures in IRAF, and the
resulting spectrum covers $\approx 3320-8530$ \AA.  As shown in
Figure~\ref{fig:lcs_spec}, the spectrum exhibits broad absorption
features at about $4500$, $5650$, $6150$, and $6800$ \AA,
corresponding to features seen in previous ULSNe
\citep{qaw+07,psb+10,ccs+11,qkk+11}.  Accounting for an expansion
velocity of about $-15,000$ km s$^{-1}$ \citep{ccs+11}, the resulting
redshift is $z\approx 1.55$.  At this redshift, emission is detected
to $\lambda_r\approx 1300$ \AA.

\begin{figure}
\centering
\includegraphics[angle=0,height=2.80in]{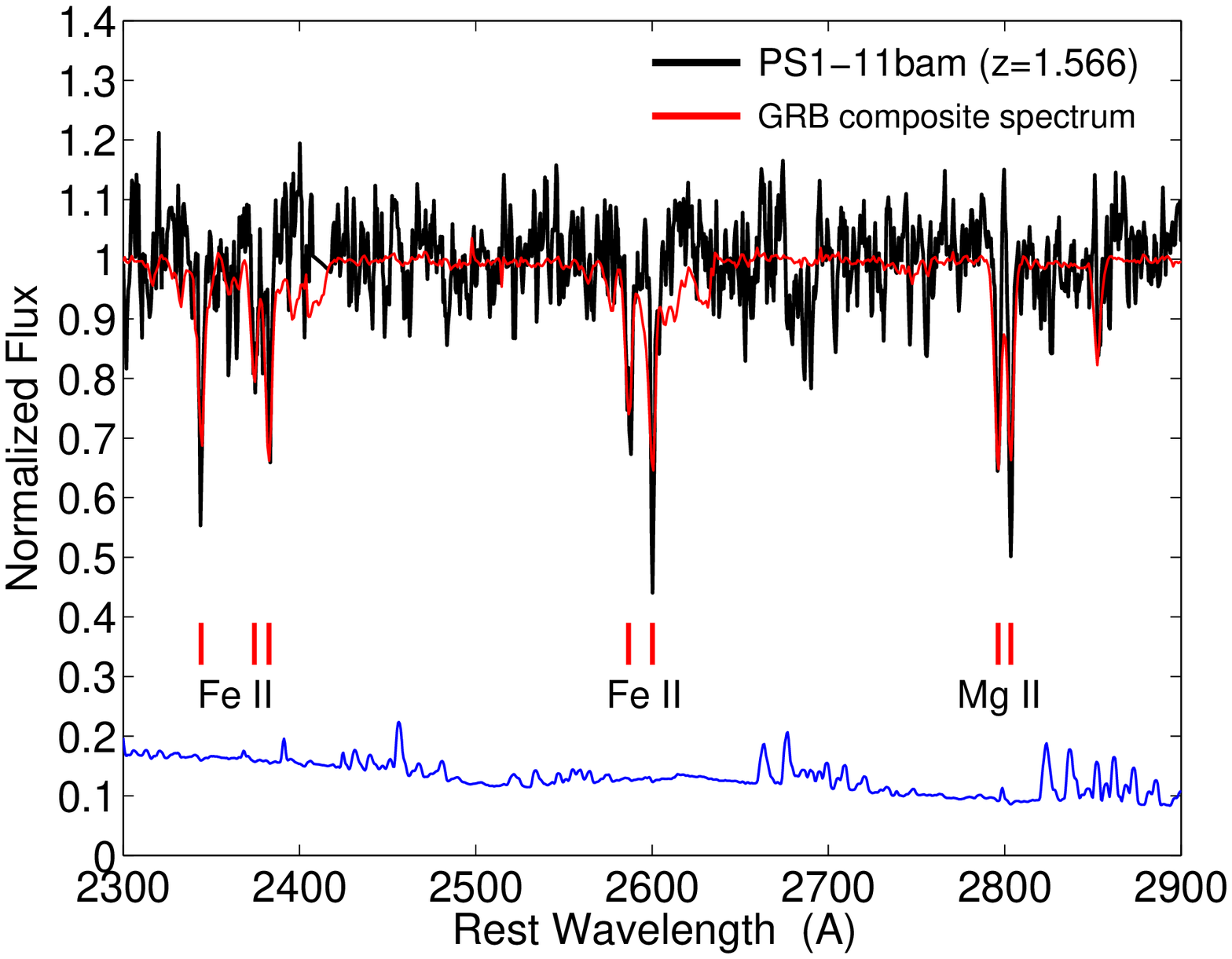}\hfill
\includegraphics[angle=0,height=2.73in]{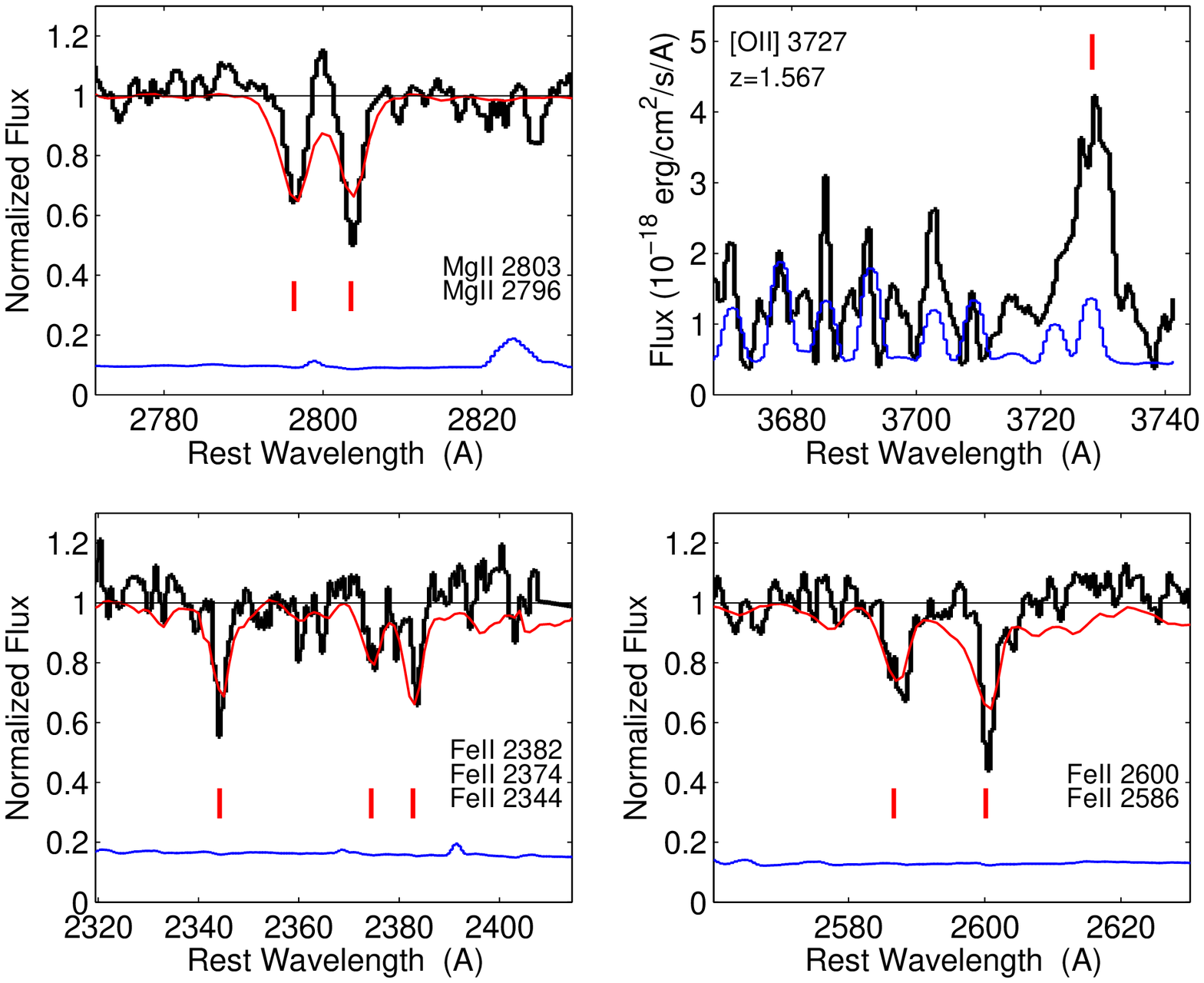}
\caption{{\it Left:} Portion of the Gemini spectrum of PS1-11bam from
December 5 containing several interstellar absorption features of
\ion{Fe}{2} and \ion{Mg}{2} at $z=1.566$ (black).  The error spectrum
is shown in blue.  For comparison we plot the GRB composite spectrum
of \citet{cfp+11}.  {\it Right:} A zoom-in on the relevant \ion{Fe}{2}
and \ion{Mg}{2} lines demonstrates the similarity to GRB absorption
spectra.  Also shown is the [\ion{O}{2}]$\lambda 3727$ emission line
at $z=1.567$ from the January 1 Gemini spectrum.  
\label{fig:absspec}}
\end{figure}

Following this identification we obtained additional spectra with the
Gemini Multi-Object Spectrograph (GMOS; \citealt{hja+04}) on the
Gemini North 8-m telescope on 2011 December 5 and 2012 January 1 UT.
The observations consisted of $2\times 1500$ s exposures with the R400
grating (December 5), $2\times 1200$ s with the B600 grating (January
1), and $2\times 1050$ s with the R400 grating (January 1) all taken
at airmass of $1.1-1.2$ and with a $1''$ slit aligned at the
parallactic angle.  The spectra were processed using the {\tt gemini}
package in IRAF, and cover $4800-9050$ \AA\ and $3850-9600$ \AA,
respectively, with a resolution of about 7 \AA.  In addition to the
broad SN features, the first Gemini spectrum reveals narrow absorption
lines of \ion{Fe}{2} and \ion{Mg}{2} at a common redshift of
$z=1.5657\pm 0.0003$ (Table~\ref{tab:lines} and
Figure~\ref{fig:absspec}).  The second Gemini spectrum further reveals
a narrow [\ion{O}{2}]$\lambda 3727$ emission line at $z=1.567\pm
0.001$ (Figure~\ref{fig:absspec}), confirming that the absorption
features arise in the host galaxy of PS1-11bam, with a potential
velocity offset of $-150\pm 80$ km s$^{-1}$.  The emission/absorption
redshift also validates our identification of the broad SN features.

The detailed properties of PS1-11bam will be discussed in a separate
paper (Lunnan et al.~in prep.); here we focus on the interstellar
absorption and the properties of the associated host galaxy.

\section{Interstellar Absorption and Comparison to Quasars and
Gamma-Ray Burst Absorbers} 
\label{sec:abs}

The detected interstellar \ion{Fe}{2} and \ion{Mg}{2} features are
shown in Figure~\ref{fig:absspec} and summarized in
Table~\ref{tab:lines}.  The rest-frame equivalent widths of the
strongest lines are $W_r({\rm Fe\,II}\lambda 2600)=1.2\pm 0.1$ \AA\
and $W_r({\rm Mg\,II}\lambda 2803)=1.3\pm 0.2$ \AA.  We use the
\ion{Mg}{2}$\lambda 2803$ line since the \ion{Mg}{2}$\lambda 2796$
line is mildly contaminated by an atmospheric sky line.  The
equivalent widths allow us to place a lower limit on the ion column
densities, assuming the optically thin regime of the curve-of-growth,
$N\gtrsim 1.13\times 10^{20}\,{\rm cm}^{-2}\,(W_r/f \lambda_r^2)$;
here $W_r$ and $\lambda_r$ are rest-frame values in units of \AA\ and
$f$ is the oscillator strength.  This is a lower limit due to likely
line saturation and the low spectral resolution.  From the strong
lines we find ${\rm log} N({\rm Fe\,II})\gtrsim 13.9$ and ${\rm log}N
({\rm Mg\,II})\gtrsim 13.8$, although the weakest oscillator strength
line (\ion{Fe}{2}$\lambda 2374$) gives ${\rm log}N({\rm Fe\,II})
\gtrsim 14.7$.  We note that the redshift of PS1-11bam is not large
enough for a detection of Ly$\alpha$, and we therefore cannot
determine the gas-phase metallicity.

\begin{figure}
\centering
\includegraphics[angle=0,width=6.4in]{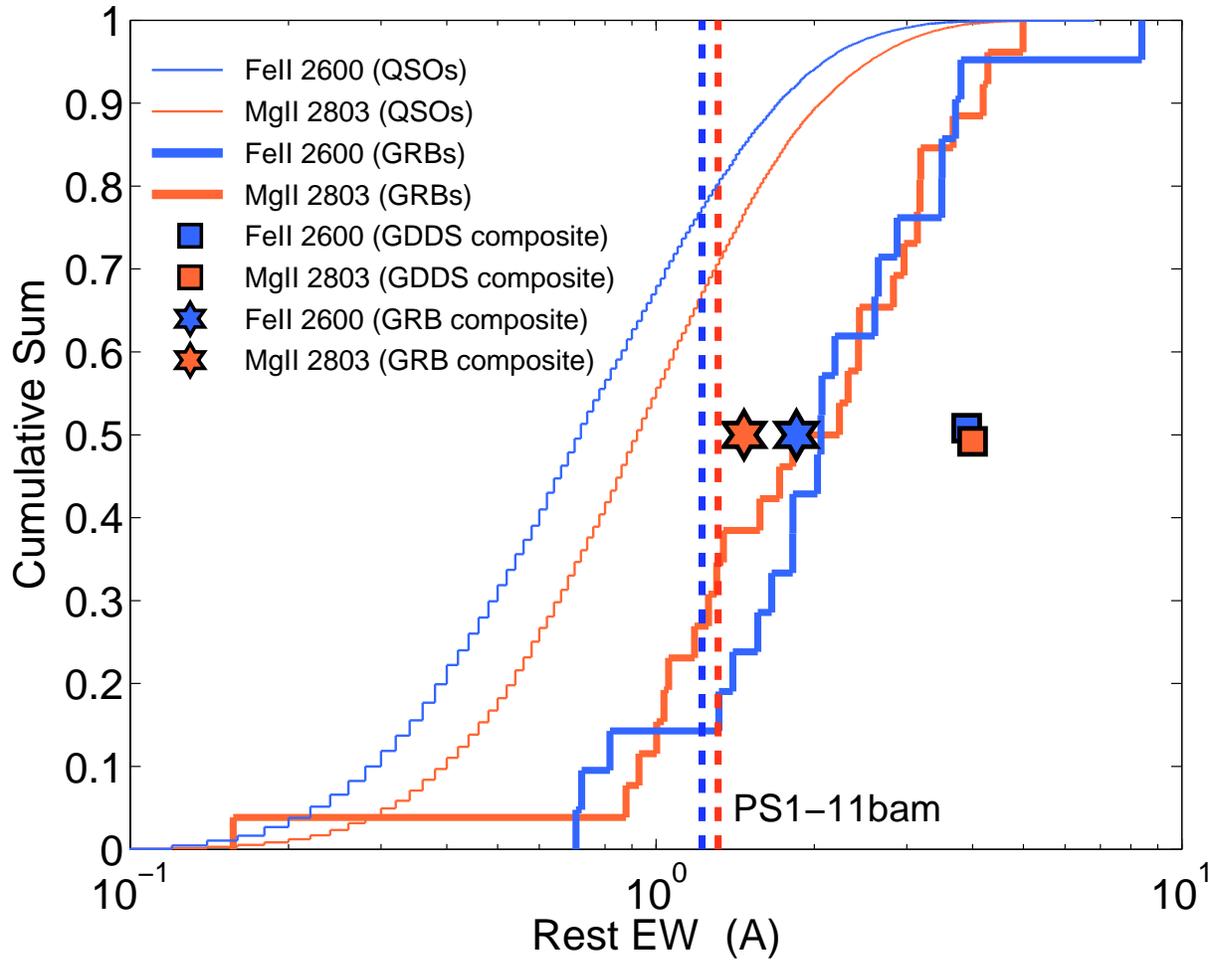}
\caption{Rest-frame equivalent widths of \ion{Mg}{2}$\lambda 2803$
(orange dashed vertical line) and \ion{Fe}{2}$\lambda 2600$ (blue
dashed vertical line) for PS1-11bam.  Also shown are the equivalent
width distributions for intervening systems at $z\approx 0.4-2.3$ from
SDSS quasar absorption spectra (thin lines; \citealt{qnt+11}),
intrinsic absorbers from GRB spectra (thick lines; \citealt{fjp+09}),
the values from a GRB composite spectrum (hexagrams; \citet{cfp+11}),
and the values from a stack of 13 star forming galaxies at $z\approx
1.3-2$ from the Gemini Deep Deep Survey (squares, offset vertically
for clarity; \citealt{sga+04}).
\label{fig:mgii}}
\end{figure}

In Figure~\ref{fig:mgii} we compare the \ion{Fe}{2}$\lambda 2600$ and
\ion{Mg}{2}$\lambda 2803$ equivalent widths to those measured for
quasar intervening systems at $z\approx 0.4-2.3$ from the Sloan
Digital Sky Survey \citep{qnt+11}, and for intrinsic absorbers (i.e.,
host galaxies) from GRB spectra \citep{fjp+09}.  We find that the
equivalent widths for PS1-11bam are in the top quartile of the quasar
absorption system distribution, but are in the bottom quartile of the
GRB intrinsic absorbers.  In comparison to the composite GRB spectrum
of \citet{cfp+11}, the \ion{Mg}{2}$\lambda 2803$ equivalent width is
comparable, with 1.3 \AA\ for PS1-11bam and 1.5 \AA\ for the GRB
composite.  The \ion{Fe}{2}$\lambda 2600$ equivalent width is lower,
with 1.2 \AA\ for PS1-11bam and 1.85 \AA\ for the GRB composite.
However, we note that the \ion{Fe}{2}$\lambda 2600$ line in the GRB
composite spectrum is blended with the fine-structure line
\ion{Fe}{2}*$\lambda 2599$, so the actual equivalent width is
$\lesssim 1.85$ \AA.  The fine-structure line is due to UV pumping by
the early afterglow intense radiation field \citep{dcp+06,vls+07},
which we do not expect\footnotemark\footnotetext{Fine-structure lines
may be excited in the local circumstellar medium in the case of
strongly-interacting SNe (e.g., \citealt{brm+00}).} in the case of
PS1-11bam.  Finally, we find that the equivalent widths for PS1-11bam
are about 3 times smaller than the values measured from a composite
spectrum of 13 star forming galaxies at a mean redshift of $z\approx
1.6$ from the Gemini Deep Deep Survey (GDDS; \citealt{sga+04}).  For
subsequent comparison with the host of PS1-bam (\S\ref{sec:host}), we
note that the GDDS galaxies have a mean rest-frame UV absolute
magnitude of $M_{2000}\approx -20.3$ AB mag.

The larger equivalent widths for the PS1-11bam absorber compared to
the bulk of quasar intervening systems, and the similarity to the GRB
composite spectrum, which is mainly composed of damped Ly$\alpha$
absorbers (DLAs; ${\rm log}N({\rm H\,I})\gtrsim 20.3$), suggests that
PS1-11bam would also uncover a DLA if we could measure its Ly$\alpha$
column density.  Indeed, quasar studies suggest that about $80\%$ of
all absorbers with \ion{Mg}{2}$\lambda 2796\,\gtrsim 1.3$ \AA, and
about $50\%$ of absorbers with \ion{Fe}{2}$\lambda 2600\,\gtrsim 1.2$
\AA\ are DLAs \citep{rtn06}.

\section{Host Galaxy Properties}
\label{sec:host}

\begin{figure}
\centering
\includegraphics[angle=0,height=2.03in]{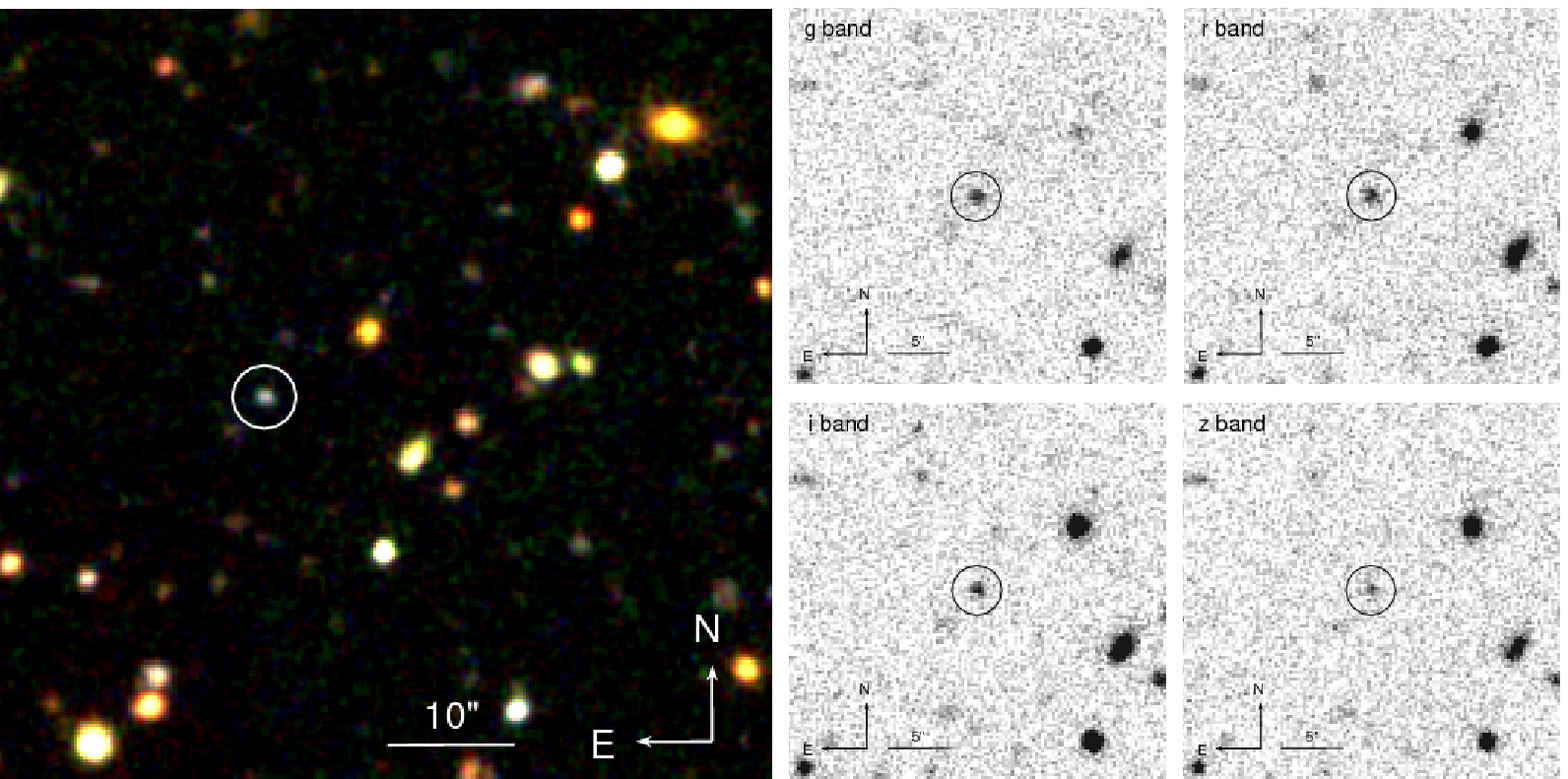}\hfill
\includegraphics[angle=0,height=2.08in]{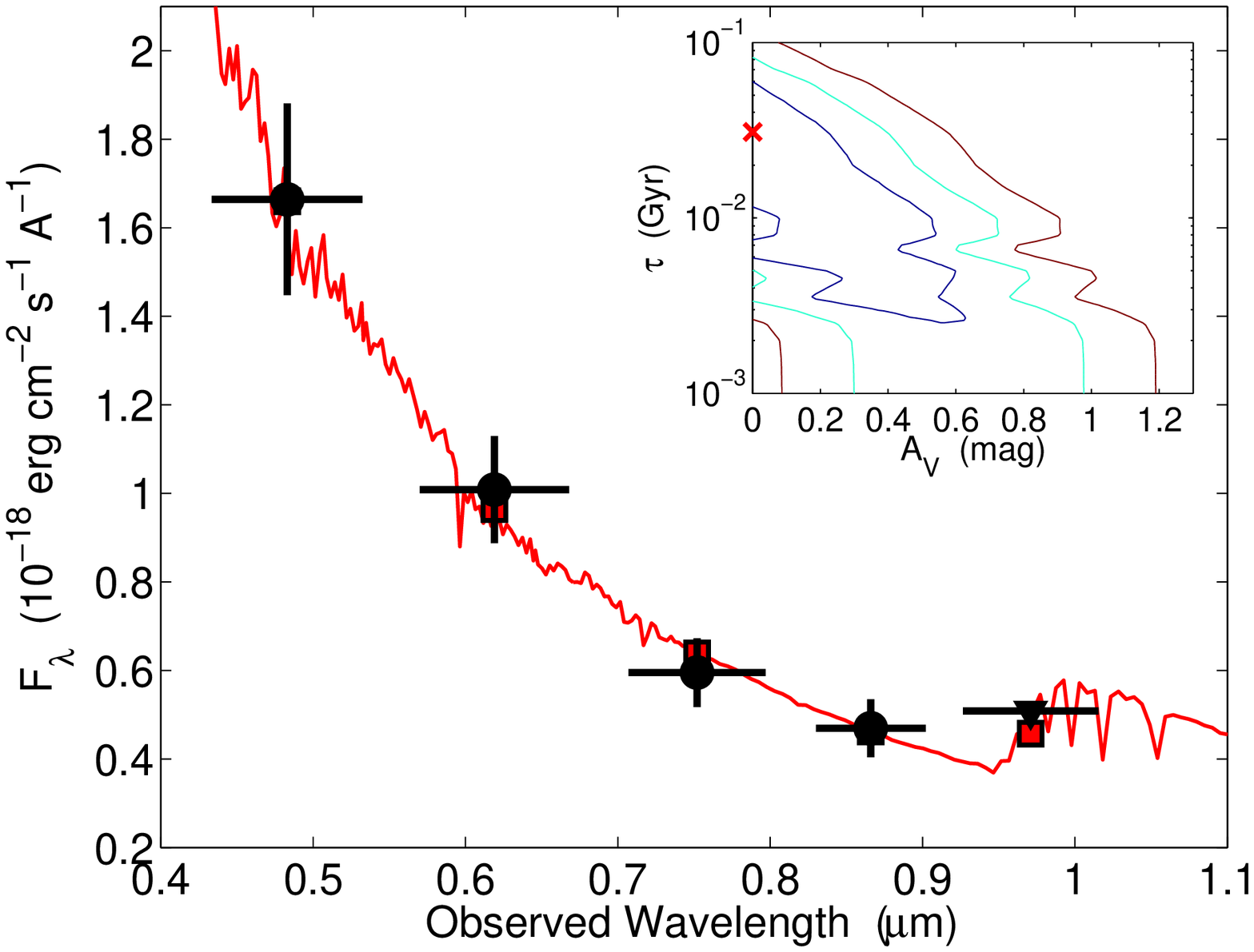}
\caption{{\it Left:} PS1/MDS pre-explosion images of the host galaxy
of PS1-11bam in \gps\rps\ips\zps, with a wider $gri$ color-composite
image demonstrating the blue colors of the host relative to nearby
field galaxies.  {\it Right:} Host galaxy SED (black; upper limit in
\yps), along with the best-fit \citet{mar05} model for $Z=0.5$
Z$_\odot$, which has $\tau_*\approx 30$ Myr, $M_*\approx 2\times 10^9$
M$_\odot$, and $A_V^{\rm host}\lesssim 0.5$ mag.  The inset shows the
$\tau_*$ versus $A_V^{\rm host}$ confidence regions with contours
marking $1\sigma$, $2\sigma$, and $3\sigma$.
\label{fig:host}}
\end{figure}

We detect the host galaxy of PS1-11bam in pre-explosion stacks of the
PS1/MDS data with apparent magnitudes\footnotemark\footnotetext{These
values are corrected for Galactic extinction of $E(B-V)\approx 0.026$
mag \citep{sfd98}.} of \gps\,$=23.62\pm 0.13$, \rps\,$=23.62\pm 0.12$,
\ips\,$=23.78\pm 0.13$, \zps\,$=23.73\pm 0.14$, and \yps\,$\gtrsim
23.4$ ($3\sigma$); see Figure~\ref{fig:host}.  The blue colors of the
galaxy are indicative of a young stellar population.  The host SED
covers rest-frame wavelengths of about $1900-3800$ \AA, with the
\yps-band limit constraining the 4000\AA/Balmer break
(Figure~\ref{fig:host}).  We fit the SED with the \citet{mar05}
evolutionary stellar population synthesis models, using a Salpeter
initial mass function and a red horizontal branch morphology, with the
stellar population age ($\tau_*$) and stellar mass ($M_*$) as free
parameters.  For a metallicity range of $0.05-1$ Z$_\odot$ we find
$\tau_*\approx 15-45$ Myr and $M_*\approx (1.1-2.6)\times 10^9$
M$_\odot$ ($\chi^2=1.2$ for $3$ degrees of freedom;
Figure~\ref{fig:host}); lower metallicity leads to older ages and
larger stellar masses.  We also find that $A_V^{\rm host}\lesssim 0.5$
mag for an assumed LMC extinction curve.

From the observed \gps-band we infer a rest-frame UV luminosity of
$L_{\rm\nu,UV}\approx 8\times 10^{28}$ erg s$^{-1}$ Hz$^{-1}$,
corresponding to a star formation rate of ${\rm SFR}\approx 12$
M$_\odot$ yr$^{-1}$ \citep{ken98}.  The upper bound on the extinction
($A_{1900}^{\rm host}\approx 1.3$ mag) indicates that ${\rm SFR}
\lesssim 40$ M$_\odot$ yr$^{-1}$.  From the [\ion{O}{2}]$\lambda 3727$
line flux, $F({\rm [O\,II]})\approx 7\times 10^{-17}$ erg cm$^{-2}$
s$^{-1}$, we find ${\rm SFR}\approx 6$ M$_\odot$ yr$^{-1}$
\citep{ken98}.  The star formation rate is substantially larger than
for previous ULSNe hosts, with ${\rm SFR}\approx 0.1-1$ M$_\odot$
yr$^{-1}$ \citep{ccs+11,nsg+11}.  The young stellar population age,
low stellar mass, and appreciable star formation rate are reminiscent
of long GRB host galaxies, for which $\langle\tau_*\rangle\sim 60$
Myr, $\langle M_*\rangle\sim 1.5\times 10^9$ M$_\odot$, and
$\langle{\rm SFR} \rangle\sim 10$ M$_\odot$ yr$^{-1}$
\citep{chg04,sgl09,lb10}.

The rest-frame UV absolute magnitude of the host galaxy of PS1-11bam,
$M_{2000}\approx -20.7$ AB mag, is brighter than the mean of the GDDS
composite spectrum, $M_{2000}\approx -20.3$ AB mag \citep{sga+04}.
This may appear puzzling given the much larger \ion{Fe}{2}$\lambda
2600$ and \ion{Mg}{2}$\lambda 2803$ equivalent widths in the GDDS
composite compared to PS1-11bam, but we note that the GDDS galaxies
are $K$-band (i.e., stellar mass) selected, while the host of
PS1-11bam was selected through a massive star explosion (i.e., star
formation).  As a result, despite its larger UV luminosity, the host
of PS1-11bam is most likely a lower stellar mass system than the mean
GDDS galaxy.

\section{Discussion and Conclusions}
\label{sec:conc}

We presented the discovery of the ULSN PS1-11bam, and follow-up
spectroscopy that reveals broad SN features and interstellar
absorption from \ion{Fe}{2} and \ion{Mg}{2}.  The ISM line equivalent
widths are larger than the median of the quasar intervening absorber
sample, but smaller than the GRB intrinsic absorber sample.  They are
substantially lower compared to a stack of GDDS galaxies at $z\approx
1.6$.  Although the potential of ULSNe as probes of distant galaxies
has been proposed previously \citep{qkk+11}, this is the first direct
demonstration of this approach at a cosmologically interesting
redshift that overlaps with the GRB and quasar samples.  This is also
the redshift range at which direct galaxy metallicity measurements
become progressively more challenging.

We also detect the host galaxy of PS1-11bam in pre-explosion images,
and find that it exhibits a substantial star formation rate, a young
stellar population age, and a low stellar mass, similar to the long
GRB host sample.  The star formation rate is substantially larger than
in previous ULSN hosts \citep{ccs+11,nsg+11}, indicating that they
likely span a wide range of properties.

The spectrum of PS1-11bam demonstrates that ISM measurements using
ULSNe are both promising and challenging.  First, the luminosities are
about a factor of $10-100$ times lower than for GRB afterglows, but
this can be partially overcome with repeated observations during the
broad peak, and in the future with 30-m class telescopes.  Second,
with a typical blackbody temperature of $\sim 2\times 10^4$ K, the
continuum flux at $\lambda_r\sim 1200$ \AA, required for absorption
measurements of Ly$\alpha$, is about 5 times lower than at the peak of
the SED.  However, our observations of PS1-11bam demonstrate that
continuum emission is indeed detectable to at least $\lambda_r\sim
1300$ \AA.  Third, any broad SN features near $\lambda_r\sim 1200$
\AA\ may complicate measurements of Ly$\alpha$ absorption; existing
data do not extend sufficiently blueward to assess this point.
Fourth, since circumstellar (CSM) interaction has been invoked to
explain the large luminosities of ULSNe \citep{ci11}, absorption in
the CSM may complicate measurements of ISM features.  However, CSM
interaction is generally accompanied by strong emission or P Cygni
line profiles, which are also time variable (e.g., \citealt{brm+00}).
This is not seen in PS1-11bam, and it may serve as a discriminant
between ISM and CSM absorption for future events.  Finally, our
observations demonstrate that at present PS1 is uniquely capable of
discovering UV-bright ULSNe to $z\sim 2$, thanks to its superior red
sensitivity and use of \ips\zps-band filters, and that follow-up
spectroscopy with 8-m class telescopes can probe ISM absorption
features; the upcoming Dark Energy Survey and PS2 will have the same
potential.  With the even greater depth of LSST, in conjunction with
follow-up spectroscopy from 30-m class telescopes, ULSNe will
ultimately probe galaxies to $z\sim 4$.

\acknowledgements PS1 has been made possible through contributions of
the Institute for Astronomy, the University of Hawaii, the Pan-STARRS1
Project Office, the Max-Planck Society and its participating
institutes, the Max Planck Institute for Astronomy, Heidelberg and the
Max Planck Institute for Extraterrestrial Physics, Garching, The Johns
Hopkins University, Durham University, the University of Edinburgh,
Queen's University Belfast, the Harvard-Smithsonian Center for
Astrophysics, and the Las Cumbres Observatory Global Telescope
Network, Incorporated, the National Central University of Taiwan, and
the National Aeronautics and Space Administration under Grant
NNX08AR22G issued through the Planetary Science Division of the NASA
Science Mission Directorate.  This work is based in part on
observations obtained at the Gemini Observatory (Program GN-2011B-Q-3;
PI: Berger), which is operated by the Association of Universities for
Research in Astronomy, Inc., under a cooperative agreement with the
NSF on behalf of the Gemini partnership: the National Science
Foundation (United States), the Science and Technology Facilities
Council (United Kingdom), the National Research Council (Canada),
CONICYT (Chile), the Australian Research Council (Australia),
Ministério da Ciência, Tecnologia e Inovação (Brazil) and Ministerio
de Ciencia, Tecnología e Innovación Productiva (Argentina).
Observations were also obtained at the MMT Observatory, a joint
facility of the Smithsonian Institution and the University of Arizona.
Some of the computations in this paper were run on the Odyssey cluster
supported by the FAS Science Division Research Computing Group at
Harvard University.


\clearpage
\begin{deluxetable}{cccccc}
\tablecolumns{6}
\tabcolsep0.5in\footnotesize
\tablewidth{0pc}
\tablecaption{ISM Absorption Lines in PS1-11bam
\label{tab:lines}}
\tablehead {
\colhead {$\lambda_{\rm obs}$} &
\colhead {Line}                &
\colhead {$f_{ij}$}            &
\colhead {$z$}                 &
\colhead {$W_r$}               &
\colhead {${\rm log}\,N$}      \\
\colhead {(\AA)}               &
\colhead {(\AA)}               &
\colhead {}                    &
\colhead {}                    &
\colhead {(\AA)}               &
\colhead {(cm$^{-2}$)}                       
}
\startdata
6014.29 & \ion{Fe}{2} 2344.214 & $0.114$ & 1.5656 & $1.0\pm 0.2$ & 14.3 \\
6092.81 & \ion{Fe}{2} 2374.461 & $0.031$ & 1.5660 & $0.7\pm 0.2$ & 14.7 \\
6114.45 & \ion{Fe}{2} 2382.765 & $0.320$ & 1.5661 & $0.7\pm 0.2$ & 13.6 \\
6638.80 & \ion{Fe}{2} 2586.650 & $0.069$ & 1.5666 & $0.8\pm 0.2$ & 14.3 \\
6672.35 & \ion{Fe}{2} 2600.173 & $0.239$ & 1.5661 & $1.2\pm 0.1$ & 13.9 \\
7174.50 & \ion{Mg}{2} 2796.352 & $0.612$ & 1.5657 & $1.0\pm 0.2$ & 13.4 \\
7192.93 & \ion{Mg}{2} 2803.531 & $0.305$ & 1.5657 & $1.3\pm 0.2$ & 13.8
\enddata
\end{deluxetable}

\clearpage

\clearpage

\clearpage

\clearpage

\end{document}